\documentclass[twocolumn,showpacs,preprintnumbers,superscriptaddress]{revtex4}
\usepackage{graphicx}
\usepackage{dcolumn}
\usepackage{bm}
\usepackage{color}
\usepackage{amsmath}
\usepackage{amssymb}
\usepackage{amsfonts}
\usepackage{esint}
\usepackage{times}
\usepackage{url}

\begin{document}
\title{Two-photon driven Kerr resonator for quantum annealing with three-dimensional circuit QED}
\author{Peng Zhao}
\affiliation{National Laboratory of Solid State Microstructures, \\
School of Physics, Nanjing University, Nanjing 210093, China}
\author{Zhenchuan Jin}
\affiliation{National Laboratory of Solid State Microstructures, \\
School of Physics, Nanjing University, Nanjing 210093, China}
\author{Peng Xu}
\affiliation{National Laboratory of Solid State Microstructures, \\
School of Physics, Nanjing University, Nanjing 210093, China}
\author{Xinsheng Tan}
\affiliation{National Laboratory of Solid State Microstructures, \\
School of Physics, Nanjing University, Nanjing 210093, China}
\author{Haifeng Yu}
\email{hfyu@nju.edu.cn}
\affiliation{National Laboratory of Solid State Microstructures, \\
School of Physics, Nanjing University, Nanjing 210093, China}
\affiliation{Synergetic Innovation Center of Quantum Information $\&$ Quantum Physics, \\
University of Science and Technology of China, Hefei, Anhui 230026, China}
\author{Yang Yu}
\affiliation{National Laboratory of Solid State Microstructures, \\
School of Physics, Nanjing University, Nanjing 210093, China}
\affiliation{Synergetic Innovation Center of Quantum Information $\&$ Quantum Physics, \\
University of Science and Technology of China, Hefei, Anhui 230026, China}
\date{\today}

\begin{abstract}
We propose a realizable circuit QED architecture for engineering states
of a superconducting resonator off-resonantly coupled to an ancillary
superconducting qubit. The qubit-resonator dispersive
interaction together with a microwave drive applied to the qubit gives rise to
a Kerr resonator with two-photon driving that enables us to efficiently
engineer the quantum state of the resonator such as generation of the
Schr\"{o}dinger cat states for resonator-based universal quantum
computation. Moreover, the presented architecture is easily scalable for
solving optimization problem mapped into the Ising spin glass model,
and thus served as a platform for quantum annealing. Although various scalable
architecture with superconducting qubits have been proposed for
realizing quantum annealer, the existing annealers are currently limited
to the coherent time of the qubits. Here, based on the protocol for
realizing two-photon driven Kerr resonator in three-dimensional circuit QED
(3D cQED), we propose a flexible and scalable hardware for
implementing quantum annealer that combines the advantage of the long coherence
times attainable in 3D cQED and the recently proposed resonator based
Lechner-Hauke-Zoller (LHZ) scheme. In the proposed resonator based LHZ
annealer, each spin is encoded in the subspace formed by two coherent
state of 3D microwave superconducting resonator with opposite phase, and thus
the fully-connected Ising model is mapped onto the network of the resonator with
local tunable three-resonator interaction. This hardware architecture provides a
promising physical platform for realizing quantum annealer with improved coherence.
\end{abstract}

\maketitle

\section{Introduction}
The parametrically driven anharmonic oscillator, which is usually modeled
by Kerr resonator with two-photon driving \cite{R1,R2}, has
been shown to display rich physics and thus has been studied extensively
\cite{R3,R4,R5,R6,R7,R8,R9,R10,R11}. In a system consisting of a Kerr resonator
with two-photon driving operated in the quantum regime, where the Kerr
nonlinearity is stronger than the photon decay rate, various schemes have been
proposed for engineering the quantum state of the resonator \cite{R4,R6,R12,R13,R14}.
Among these efforts, preparing the Schr\"{o}dinger cat state, i.e., superpositions
of two large coherent states with opposite phases, has been extensively investigated
for the development of quantum metrology \cite{R15} and quantum information processing \cite{R15,R16}.
Moreover, driven by the pursuit towards practical quantum information processing,
the quantum annealing \cite{R17} was proposed as a quantum enhanced optimizer that aims to
efficiently solve Ising problems \cite{R18,R19}, and this resonator
system has attracted increasing attention owing to its coupled network offering an
new paradigm for quantum annealer \cite{R12,R20,R21,R22}. In these paradigm,
the quantum information is encoded and protected in
the continuous variable system, i.e., resonator, and a series of theoretical
studies have shown the robustness of these paradigm with respect to the dissipation and
noise \cite{R21,R22}.

Typically, the Kerr nonlinearity can be induced by inserting a nonlinear
medium (Kerr medium) in a resonator. However, the induced Kerr nonlinearity
is usually smaller than the dissipation rate in optical and mechanical
systems, hindering the study of the quantum regime of this system.
Recent advances in artificial solid-state system, especially in the
cQED system, provides an alternative approach to easily get
access to this fascinating quantum regimes \cite{R23}. In the context of cQED,
the Kerr resonator is realized by using a superconducting resonator
with an embedded Josephson junction \cite{R24,R25} which is almost an ideal
non-dissipative nonlinear element \cite{R26}. Therefore, this gives rise to two main approaches to
realize the two-photon driven Kerr resonator in the cQED architecture.
One approach to realizing the system is to use a superconducting coplanar resonator
terminated by a flux-pumped SQUID \cite{R27,R28,R29}. Alternatively, a recent
experimental work has shown a nonlinear driven-dissipation approach by using
a 3D microwave resonator coupled to fixed-frequency transmon qubit and by an
external microwave drive applied to the qubit at the resonator frequency \cite{R30}.
For the latter approach, remarkably, the 3D microwave resonator with high
quality factors has been experimentally demonstrated \cite{R31}, enabling
storage times approaching seconds \cite{R32} and thus favoring the quantum
information processing based on the 3D microwave superconducting resonator \cite{R16,R33}.
However, it is noted that this approach is implemented by using the two-photon driven-dissipation
process where the engineered nonlinear (two-photon) decay rate should be
significantly larger than the single-photon decay rate of the cavity,
which still remains important challenges to be overcome \cite{R30}.
This current limitation suggests investigating another relatively
easily realizable approach that also exploits the large coherence
times of the 3D microwave superconducting resonator.

In this work we proposed an experimentally feasible protocol for realizing
two-photon driven Kerr resonator in a cQED architecture, which, in principle,
is compatible with the 3D cQED architecture with high coherence. The Kerr
resonator with two-photon driving, which lie at its heart, is realized via
off-resonantly coupling a resonator to an ancillary microwave-driven
superconducting qubit. In our setting, the superconducting qubit is treated as 
a quantum three-level system that is initially in its ground state, and remains unexcited in the
whole process, minimizing the effect of the qubit decoherence. Moreover, based
on our protocol for realizing two-photon driven Kerr resonator, we apply a
combination of the recently proposed resonator based LHZ annealer \cite{R21}
and the 3D cQED architecture that allows a flexible and scalable hardware
architecture with long coherence times for LHZ annealer. In our setting,
each spin in the LHZ represented Ising problem is encoded in the degenerate
ground subspace of the two-photon driven Kerr resonator formed by two coherent
states of opposite phases, and the four-body constraints, which are decomposed
into two three-body constraints, are physical implemented with the use of a
transmon qubit mediated tunable three-resonator interactions.

\section{the system and Hamiltonian}
\begin{figure}[tbp]
\begin{center}
\includegraphics[width=8.0cm,height=6cm]{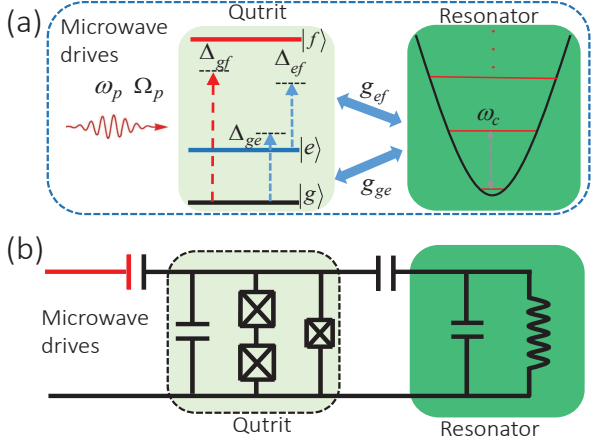}
\end{center}
\caption{(Color online) (a) General proposal for realizing the two-photon driven
Kerr resonator, where the transitions $|g\rangle\leftrightarrow|e\rangle$
and $|e\rangle\leftrightarrow|f\rangle$ of the quantum three-level system (qutrit)
are dispersively coupled with the resonator, while the $|g\rangle\leftrightarrow|f\rangle$
transition is driven off-resonantly by a coherent microwave drive of frequency at the
nearly twice the resonator frequency. (b) Circuit diagram of the proposed circuit-QED system
for realizing the proposal, where the superconducting resonator is capacitively coupled to a
superconducting flux qubit that is driven by a coherent microwave drive. }
\end{figure}

In order to construct a circuit-QED platform for engineering quantum
state of resonator, and further implementing the quantum annealing,
here we introduce two major ingredients, two-photon driven Kerr resonator,
which allows us to create non-classical state of resonator such as
Schr\"{o}dinger cat states, and the tunable resonance three-resonator interaction.
Meanwhile, we also aim to exploit the long coherence of the 3D superconducting
resonator, thus the presented platform should be compatible with the 3D architecture.
With these aims in mind, in the following discussion: (i) We propose a scheme to realize
the two-photon driven Kerr resonator in a qubit-resonator system, where
the resonator is coupled to a superconducting qubit and the qubit is driven
by a coherent microwave drive applied at well-chosen frequency, as
shown in Fig.$\,$1. (ii) We show that a tunable resonance three-resonator
interaction can be induced by coupling resonators to a transmon qubit,
where the qubit mode decouples form these resonator modes and mediates the
three-resonator interaction by applying a suitable pump mode on it, as
shown in Fig.$\,$3. In doing so, we present a thorough description of
the two ingredients for the coupled network of two-photon driven Kerr
resonators. We now turn to present the quantitatively derivation.

\subsection{Kerr-nonlinear resonator with two-photon driving}

\begin{figure*}[tbp]
\begin{center}
\includegraphics[width=16cm,height=7cm]{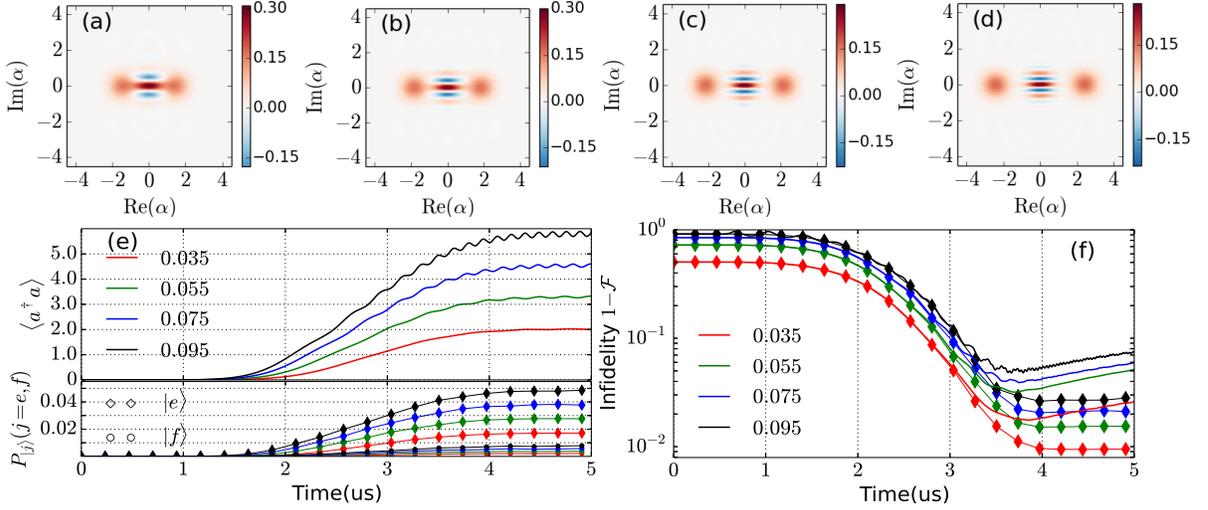}
\end{center}
\caption{(Color online) Quantum adiabatic evolution of the system with
a microwave drive $\Omega_{p}(t)=\Omega_{p}[1-e^{-(t/\tau)^{4}}]$,
where system is initially in its ground state, i.e., the vacuum
state $|0\rangle$ for the resonator, and the $|g\rangle$ for the qutrit.
The whole time of the adiabatic evolution $T=5\,\mu$s,
and $\tau=3\,\mu$s. (a)-(d) Wigner function for the resonator at the
end of the adiabatic evolution for $\Omega_{p}/2\pi=(0.035,0.055,0.075,0.095)$ GHz,
respectively. (e) Time evolution of the average photon number
$\langle a^{\dagger}a\rangle$ and population leakage to the
excited state $(|e\rangle, |f\rangle)$ of the qutrit during the
adiabatic evolution. (f) Time dependent of the fidelity of the
cat state $\mathcal{F}(t)$. Parameters used in the numerical simulation are $\omega_{c}/2\pi=5.25$ GHz,
$\varepsilon_{e}/2\pi=6.25$ GHz, $\varepsilon_{f}/2\pi=10.0$ GHz,
$g_{ge}/2\pi=0.094$ GHz, $g_{ef}/2\pi=0.136$ GHz, and $g_{gf}/2\pi=0.140$ GHz.
In (f), for the numerical simulation of the dynamics under the influence
of dissipation, we use a decay rate of the resonator at $\kappa=1/500$ MHz, and the
decay rate of the qutrit with $\gamma_{ge}=\gamma_{gf}=1/1.5$ MHz and
$\gamma_{ef}=1/1.0$ MHz.}
\end{figure*}

As shown in Fig.$\,$1(a), we consider a system consisting of a resonator
coupled to a quantum three-level system (qutrit), for which the transitions
$|g\rangle\leftrightarrow|e\rangle$ and $|e\rangle\leftrightarrow|f\rangle$
of the qutrit are off-resonantly coupled with the resonator, while the
$|g\rangle\leftrightarrow|f\rangle$ transition is coupled to a coherent
microwave drive with a frequency at the nearly twice the resonator frequency.
The proposed scheme can be physically realized in a circuit QED architecture consisting
of a superconducting resonator capacitively coupled to a superconducting
qubit, which can be realized by using flux qubit \cite{R37,R38} or fluxoninum qubit \cite{R39},
as depicted in Fig.$\,$1(b). Here, we focus on the lowest three energy
levels of the superconducting qubit and thus we treat the
qubit as a three-level system (qutrit) (see Appendix A for details of the
circuit-QED system). After applying the rotating wave approximation (RWA), the
full system can be described by the Hamiltonian ($\hbar =1$, throughout the paper)
\begin{eqnarray}
\begin{aligned}
H=H_{0}+H_{I}+H_{d},
\end{aligned}
\end{eqnarray}
where
\begin{eqnarray}
\begin{aligned}
H_{0}=\omega_{c}a^{\dagger}a+\sum_{j=g,e,f}\epsilon_{j}|j\rangle\langle j|,
\end{aligned}
\end{eqnarray}
describes the free Hamiltonian of the resonator and the qutrit,
\begin{eqnarray}
\begin{aligned}
H_{I}=g_{ge}(|g\rangle\langle e|a^{\dagger}+|e\rangle\langle g|a)+g_{ef}(|e\rangle\langle f|a^{\dagger}+|f\rangle\langle e|a),
\end{aligned}
\end{eqnarray}
describes the qutrit-resonator interaction, and
\begin{eqnarray}
\begin{aligned}
H_{d}=\Omega_{p}(e^{i\omega_{p}t}|g\rangle\langle f|+e^{-i\omega_{p}t}|f\rangle\langle g|),
\end{aligned}
\end{eqnarray}
describes the time-dependent drive of the qutrit. Above, $a^{\dagger }$, $a$
are the creation and annihilation operators for the resonator of frequency
$\omega _{c}$. $\epsilon_{j}\,(j=g,e,f)$ is the transition
frequency of the qutrit from ground to excited state $|j\rangle$.
$g_{ge}$ and $g_{ef}$ denote the qutrit-resonator coupling strength, and $\Omega_{p}$
is the real amplitude of the microwave drives at frequency
$\omega _{p}\approx2\omega _{c}$. For simplicity, we define
$\epsilon_{g}=0$ in the following discussion.

We consider that our system operates in the dispersive regime, where
the qutrit is far detuned from the resonator $|\Delta _{jk}|=|(|\epsilon_{j}
-\epsilon_{k}|)-\omega _{c}|\gg g_{jk}$, and also the microwave drive $|\Delta _{d}^{\prime }|
=|\epsilon_{f}-\omega_{p}|\gg \Omega_{p}$. In this situation, the systemic
Hamiltonian $H$ can be well approximated by the effective Hamiltonian \cite{R40,R41,R42}
\begin{eqnarray}
\begin{aligned}
H_{eff}=\widetilde{\omega}_{c} a^{\dagger}a+Ka^{\dagger2}a^{2}-P(a^{\dagger 2}
e^{-i\omega_{p}t}+a^{2}e^{i\omega_{p}t})
\end{aligned}
\end{eqnarray}
where $\widetilde{\omega}_{c}=\omega_{c}+S$ is the renormalized resonator
frequency
\begin{eqnarray}
\begin{aligned}
S=-\frac{g_{ge}^{2}}{\Delta_{ge}}+\frac{g_{ge}^{4}}{\Delta_{ge}^{3}}.
\end{aligned}
\end{eqnarray}
The second term denotes the qutrit-induced self-Kerr-nonlinearity of the
resonator with
\begin{eqnarray}
\begin{aligned}
K=-\frac{g_{ge}^{2}g_{ef}^{2}}{\Delta_{ge}^{2}(\Delta_{ge}+\Delta_{ef})}+\frac{g_{ge}^{4}}{\Delta_{ge}^{3}}.
\end{aligned}
\end{eqnarray}
The last term represents a two-photon drive of amplitude
\begin{eqnarray}
\begin{aligned}
P=-\frac{g_{ge}g_{ef}\Omega_{p}}{\Delta_{ge}(\epsilon_{f}-\omega_{p})}
\end{aligned}
\end{eqnarray}
applied on the resonator at frequency $\omega_{p}=2\widetilde{\omega}_{c}$.

In deriving Eq.$\,$(5), it is worth mentioning that: (i) We have
also assumed that the qutrit is initially in its ground state. As the
qutrit-resonator system operates in the far-detuned dispersive regime,
no energy is exchanged between the field mode (resonator and the microwave field)
and the qutrit, thus the qutrit remains unexcited. Therefore,
one can eliminate the degrees of freedom of the qutrit and the effective
Hamiltonian is obtained. (ii) For a realistic implementation depicted in Fig.1(b),
we have ignored several terms due to their negligible effects, the highly
off-resonance coupling between transitions (i.e, $|g\rangle\leftrightarrow|e\rangle$,
$|e\rangle\leftrightarrow|f\rangle$) and the qutrit drives. For clarity, we have
also omitted the term describing the coupling between qutrit transition
$|g\rangle\leftrightarrow|f\rangle$ and resonator in the Hamiltonian in Eq.$\,$(1).
However, this does not alter the main result except making another contributions
to the coefficient $S$ and $K$, see Appendix A for the complete derivation.
Moreover, these terms are taken into account in the following
numerical analysis.

As a result, we have demonstrated that the qutrit-resonator system depicted
in Fig.$\,$1 can be modeled as a two-photon driven Kerr resonator. Recent
theoretical studies have shown that the schr\"{o}dinger cat state can be
generated via quantum adiabatic evolution of the system described by the
Hamiltonian Eq.$\,$(5) \cite{R12,R13}. This can be made clear by moving to a rotating
frame with respect to $\omega_{p}a^{\dagger}a/2$ such that the Hamiltonian Eq.$\,$(5)
is simplified to \cite{R13}
\begin{eqnarray}
\begin{aligned}
H&=Ka^{\dagger2}a^{2}-P(a^{\dagger 2}+a^{2})
\\&=K(a^{\dagger 2}-\frac{P}{K})(a^{2}-\frac{P}{K})-\frac{P^{2}}{K}.
\end{aligned}
\end{eqnarray}
For simplicity, we assume that $K$ and $P$ are positive for the following discussion. It is apparent
that the ground state is two-fold degenerate. The coherent states
$|\pm\alpha\rangle (\alpha=\sqrt{P/K})$, are the degenerate eigenstates
with energy $-P^{2}/K$, and thus also the schr\"{o}dinger cat states
\begin{eqnarray}
\begin{aligned}
&|\mathcal{C}_{\alpha}^{\pm}\rangle=\mathcal{N}_{\alpha}^{\pm}(|\alpha\rangle\pm|-\alpha\rangle),\,\,\,
\mathcal{N}_{\alpha}^{\pm}=\sqrt{2(1\pm e^{-2\alpha^{2}})}
\end{aligned}
\end{eqnarray}
where the $\mathcal{N}_{\alpha}^{\pm}$ is the normalizing factor, and the $\pm$
label the even- and odd cat-states, respectively. Moreover, the vacuum state $|0\rangle$ and
the one-photon Fock state $|1\rangle$ are also the ground eigenstates with even and odd
parity for the undriven case ($P=0$). As the Hamiltonian
preserves the parity, when one gradually increases the amplitude of the two-photon
drive $P(t)$, the driven system will evolve adiabatically along two paths
$|\mathcal{C}_{\alpha_{(t)}}^{\pm}\rangle$ with $\alpha_{(t)}=\sqrt{P(t)/K}$ for
the system initially prepared in the vacuum state and single-photon Fock
state, respectively \cite{R12,R13}.

To show the validity of our proposal for the two-photon driven Kerr resonator,
we present in Fig.$\,$2 the numerical analysis of the time evolution of the
qutrit-resonator system, initially in $|0,g\rangle$, with a microwave drive
$\Omega_{p}(t)=\Omega_{p}[1-e^{-(t/\tau)^{4}}]$ applied on the qutrit \cite{R13} (see Appendix B). The total
evolution time $T$ and $\tau$ are chosen to satisfy the adiabatic condition,
and we use $T=5\,\mu$s and $\tau=3\,\mu$s in this work. Parameters used in the
numerical simulation are $\omega_{c}/2\pi=5.25$ GHz, $\varepsilon_{e}/2\pi=6.25$ GHz,
$\varepsilon_{f}/2\pi=10.0$ GHz, $g_{ge}/2\pi=0.094$ GHz, $g_{ef}/2\pi=0.136$ GHz,
and $g_{gf}/2\pi=0.140$ GHz, yielding $K/2\pi\approx0.450$ MHz. For the microwave
drive with different amplitude (i.e., $\Omega_{p}/2\pi=0.035,0.055,0.075,0.095$ GHz), Figure$\,$2(a)-(d) show
the Wigner function for the resonator at the end of the adiabatic evolution, respectively,
and the resonator evolves to the even cat state $|\mathcal{C}_{\alpha}^{+}\rangle$ as
we expected. The time evolution of the average photon number in resonator is also displayed in
the upper panel of Fig.$\,$2(e), and the photon number at time $t=T$ shows good agreement
with the value calculated by $P/K$, yielding the average photon number $(2.08,3.28,4.47,5.66)$
for the four different drive amplitudes, respectively.

The lower panel of Figure$\,$2(e) shows the population leakage to the excited state
$(|e\rangle, |f\rangle)$ of the qutrit, where the diamond-markered line and circle-markered
line represent the leakage to $|e\rangle$ and $|f\rangle$, respectively, which leads
to an important limitation of our proposal for larger drive amplitudes. These
results show that for the fixed system parameter, increasing $\Omega_{p}$ will
increase the population leakage, causing undesirable entanglement of the resonator
and qutrit and adding another decay channel of the resonator. Figure$\,$2(f)
shows the time dependent of the fidelity of the cat state
$\mathcal{F}(t)= \sqrt{\langle \mathcal{C}_{\alpha}^{+}|\rho_{r}(t)|\mathcal{C}_{\alpha}^{+}\rangle}$,
in which $\alpha=\sqrt{P/K}$, and $\rho_{r}(t)$ is the
reduced density matrix of the resonator. The markered line represents
the time evolution of the fidelity without the effect of dissipation, while
the solid line corresponds to the situation by considering the effect of
photon decay and qutrit relaxation. As we expected, increasing the microwave drive
amplitude will increase the infidelity. Meanwhile, the infidelity caused by these population
leakage can be made smaller by increasing the qutrit-resonator detuning. However, we
note that increasing the qutrit-resonator detuning would entail a sacrifice of the magnitude
of the Kerr nonlinearity and the two-photon drive.

\subsection{Tunable resonant interactions among three resonators}

\begin{figure}[tbp]
\begin{center}
\includegraphics[width=8cm,height=3cm]{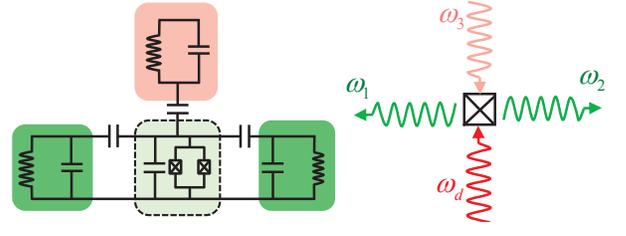}
\end{center}
\caption{(Color online) Left panel: Circuit diagram of the system consisting of
three resonators coupled to a superconducting transmon qubit, which can be
used for implementation of the tunable resonant three-body interactions. Right panel: Diagrammatic
representation of the mechanism behind the effective resonant three-body
interactions: four-wave mixing process in nonlinear element, i.e., Josephson
junction, induced by the pump at the frequency
$\omega_{d}=\omega_{1}+\omega_{2}-\omega_{3}$, where a pump photon in
combination with a resonator photon (with frequency $\omega_{3}$)
create a photon in both the other two resonators at frequencies
$\omega_{1}$ and $\omega_{2}$, respectively.}
\end{figure}
Here, we present a scheme for realizing tunable three-resonator interaction
and the interaction of interest to us is
\begin{eqnarray}
\begin{aligned}
H_{3body}=J_{123}(a^{\dag}_{1}a^{\dag}_{2}a_{3}+a_{1}a_{2}a^{\dag}_{3}),
\end{aligned}
\end{eqnarray}
where the $a^{\dag}_{j}$ and $a_{j}$ are the creation and annihilation operators for the $j$th
resonator of frequency $\omega _{j}$. As shown in Fig.$\,$3, we consider a system
consisting of three strongly detuned resonators dispersively coupled to a transmon
qubit. With the nonlinear cosine-coupling contributing from the Josephson junction
of the transmon qubit, four-wave mixing processes that conserve energy, can
happen by applying a suitable pump mode on the qubit. Similar process have been
experimentally demonstrated in 3D cQED architecture \cite{R30,R43,R43}. Here, we looks for a
four-wave mixing process, where a pump photon in combination with a resonator photon
(with frequency $\omega_{3}$) can create a photon in both the other two
resonators at frequencies $\omega_{1}$ and $\omega_{2}$, respectively, that
results the three-body interaction described by the interaction Hamiltonian
given in Eq.$\,$(11).

Now, we turn to give a quantitatively derivation of the effective Hamiltonian.
The Hamiltonian of the system depicted in Fig.$\,$3 is \cite{R30}
\begin{eqnarray}
\begin{aligned}
H=&\omega^{(0)}_{q}a^{\dag}_{q}a_{q}+\sum_{j=1}^3\omega^{(0)}_{j}a^{\dag}_{j}a_{j}-E_{J}(\cos{\varphi}+\frac{1}{2}\varphi^{2})
\\&+2\varepsilon_{p}\cos(\omega_{d}t)(a^{\dag}_{q}+a_{q}),
\end{aligned}
\end{eqnarray}
where $a^{\dag}_{m}$ and $a_{m}$ ($m=q,1,2,3$) are the creation and annihilation operators for
the $m$th mode with bare frequency $\omega^{(0)}_{m}$, and $E_{J}$ is Josephson energy
of the qubit mode. $\varphi=[\phi_{q}(a^{\dag}_{q}+a_{q})+\sum_{j=1}^3\phi_{j}(a^{\dag}_{j}+a_{j})]$
is the phase difference across the junction, and $\phi_{m}$ ($m=q,1,2,3$) is the zero-point fluctuation of flux
associated with the $m$th mode. The term in the second line of Eq.$\,$(12)
describes a pump mode with real amplitude $2\varepsilon_{p}$ applied on
the qubit mode at frequency $\omega_{d}$.

In order to see clearly how the four-wave mixing process can lead to the desired three-resonator
interaction, it is helpful to move to a displaced frame by performing
a time-dependent transformation $U(t)=e^{-\tilde{\xi}_{p}a^{\dag}_{q}+\tilde{\xi}^{\ast}_{p}a_{q}}$ with
$\tilde{\xi}_{p}=\xi_{p}e^{-i\omega_{d}t}$ and $\xi_{p}=\frac{\varepsilon_{p}}{\omega_{d}-\omega_{q}}$
on the above Hamiltonian \cite{R30}. Assuming small phase fluctuations, we can expand
the cosine up to the fourth order, and the resulting Hamiltonian after a rotating wave
approximation in the displaced frame reads (see Appendix C for more details)
\begin{eqnarray}
\begin{aligned}
H=&\sum_{j=1}^3\omega_{j}a^{\dag}_{j}a_{j}+
J_{123}(a^{\dag}_{1}a^{\dag}_{2}a_{3}e^{-i\omega_{d}t}+a_{1}a_{2}a^{\dag}_{3}e^{i\omega_{d}t})
\end{aligned}
\end{eqnarray}
where $\omega_{j}$ is the frequency for the $j$th mode including a renormalization
of the transition frequency coming from the qubit-resonator coupling and the pump mode
induced AC Stark shift. $J_{123}=-E_{J}\phi_{1}\phi_{2}\phi_{3}\phi_{q}\xi_{p}$
is the three-body coupling strength, which can be controlled by the pump drive. When
the pump frequency $\omega_{d}$ matches the detuning of the three resonators, i.e.,
$\omega_{d}=\omega_{1}+\omega_{2}-\omega_{3}$, the interaction Hamiltonian given
in Eq.$\,$(11) is obtained. Moreover, by tuning the magnitude and the phase of the
pump drive, one can realize an amplitude- and phase-tunable three-resonator
interaction, which is useful to implement the ramp protocol for quantum annealing,
as we demonstrated in the following section.

\section{resonator based Lechner-Hauke-Zoller annealer with three-body constraints}
\begin{figure}[tbp]
\begin{center}
\includegraphics[width=8.0cm,height=6cm]{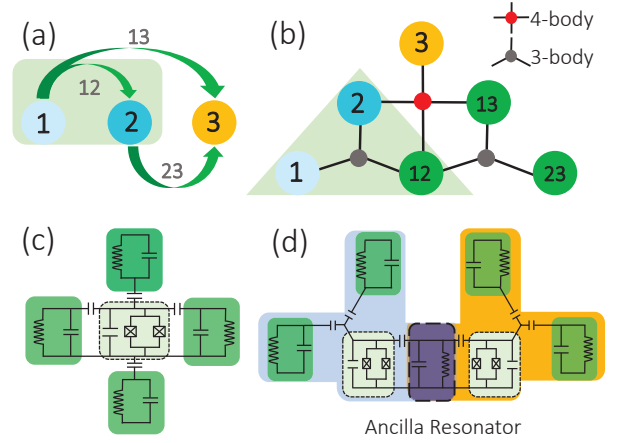}
\end{center}
\caption{(Color online) Illustration of the fully connected architecture with three-body
constraints. (a) Graph of the fully connected Ising problem with $N=3$ logical qubits.
(b) LHZ implementation of the same problem with $N_{p}=N(N+1)/2=6$ physical qubits and
$N_{C}=N(N-1)/2=3$ local constraints.
(c) Circuit-QED architecture for the physical implementation of the four-body constraints needed
in the Kerr-Resonator based LHZ scheme. (d) Decomposition of the four-body constraints
into two three-body constraints by using an ancillary resonator. }
\end{figure}

In this section, we show that our procedure to realize the two-photon driven Kerr
resonator and the tunable resonance three-resonator interaction in the circuit-QED
architecture, as demonstrated in Sec.II, could be used for the implementation of
quantum annealing with LHZ scheme.

Here, for easy reference and to set the notation, we briefly review some basic
concepts of quantum annealing and also the LHZ scheme. The quantum annealing was
proposed as a quantum enhanced optimizer that aims to efficiently solve optimization
problems. In quantum annealing, one can map an optimization problem into the all-to-all
Ising spin glass model \cite{R45}
\begin{eqnarray}
\begin{aligned}
H_{P}=&\sum_{j=1}^{N}h_{j}\sigma_{j}^{Z}+\sum_{(j<k)}\mathcal{J}_{jk}\sigma_{j}^{Z}\sigma_{k}^{Z}
\end{aligned}
\end{eqnarray}
where $\sigma_{j}^{Z}$ is the Pauli operator for the $j$th spin. The local
field $h_{j}$ and the strength of the spin-spin coupling $\mathcal{J}_{jk}$
fully define the optimization problem. The solution of the optimal problem
now amounts to finding the ground state of the Ising spin glass model
(Ising problem), and this can be achieved by executing the time-dependent
Hamiltonian
\begin{eqnarray}
\begin{aligned}
H(t)=(1-\frac{t}{T})H_{I}+(\frac{t}{T})H_{P}
\end{aligned}
\end{eqnarray}
where $H_{I}$ is the initial Hamiltonian with a trivial ground state
(e.g., $H_{I}=\sum_{j=1}^{N}b_{j}\sigma_{j}^{Z}$) and $T$ is the total evolution time.
For a system, which is governed by the Hamiltonian $H(t)$ and initially in
its ground state, evolving adiabatically, the system will stay in the instantaneous
ground state of the Hamiltonian at each time $t$. Therefore, at the end of the
evolution $t=T$, the system will stay in the ground state of $H_{P}$, which
encodes the desired solution of the optimal problem.

However, to solve a practical optimal problem mapped into the Ising model with full
connectivity, one leading physical restriction that makes the direct physical
implementation of the model becomes untractable is that interactions between
physical systems are commonly local, which favors local interaction between
spins rather than long range interaction. To get ride of this obstacle,
an embedding technique, now known as minor embedding scheme, was first introduced \cite{R46,R47}.
Recently, Lechner \textit{et al.}$\,$ \cite{R34} proposed an alternative embedding scheme
in which the full connected Ising model with $N$ logical spins [Fig.$\,$4(a)] is encoded
in $N_{P}=N(N+1)/2$ physical spins with $N_{C}=N(N-1)/2$ local constraints in a triangular
lattice, as shown in Fig.$\,$4(b) \cite{R48}. In the LHZ scheme, each physical spin
encodes the relative orientation of the corresponding pair of logical spins, i.e., the
physical spin take $|\uparrow\rangle$ and $|\downarrow\rangle$ for aligned
(i.e., $|\uparrow\uparrow\rangle$, $|\downarrow\downarrow\rangle$) and antialigned
(i.e., $|\uparrow\downarrow\rangle$, $|\downarrow\uparrow\rangle$) logical pair, respectively.
Meanwhile, the local constraints are introduced for suppressing the
redundancy of the encoding scheme. With this scheme, only local four-body terms and programmable
local fields applied on the physical spin are needed \cite{R49}. These features along with the
potential of scaling up make it attractive for practical physical implementation, and various physical
implementation scheme have been proposed \cite{R21,R35,R36,R50} and demonstrated
experimentally for small system \cite{R51}. In particular, the resonator-based
implementation proposed in Ref.$\,$[21] has been shown to be realizable and noise-resilient.
Therefore, in the following discussion, by combining the advantage of this resonator-based
implementation and the long coherence times attainable in 3D cQED, we propose a scalable
hardware for implementing resonator-based LHZ annealer with three-body constraints, which is
compatible with 3D cQED architecture.

\subsection{the Lechner-Hauke-Zoller scheme with three-body constraints}

\begin{figure}[tbp]
\begin{center}
\includegraphics[width=8cm,height=4.0cm]{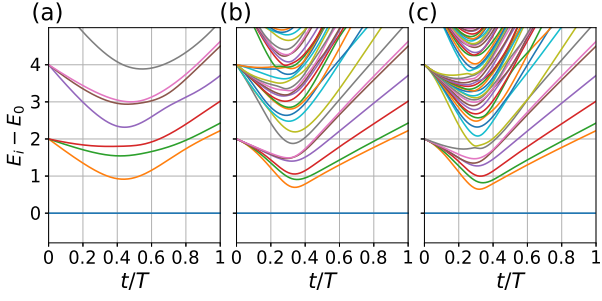}
\end{center}
\caption{(Color online) Time-dependent spectrum of an annealing process for
the Ising problem with $N=3$ logical spins. The parameters of the Ising problem
used for numerical simulation are: the local field acted on the $j$th logical
spin $h_{j}$, and the spin-spin coupling strength $\mathcal{J}_{jk}$ are random
number taken from the interval $[-\mathcal{J},\mathcal{J}]$. (a) Fictitious
directly implementation. (b) Implementation with LHZ-scheme where each four-body
constraint is realized directly by using a four-spin interaction. (c) Implementation
with LHZ-scheme where each four-body constraint is represented by two three-body
constraints that are physically realized by using three-spin interactions.
Here, $t$ is the time and $T$ is the total evolution time, and $E_{i}$ is the
eigenenergy. We use a constraint strength $C/\mathcal{J}=3$. }
\end{figure}
In the LHZ scheme, each four-body constraint can be realized directly
by using a four-spin interaction $\sigma_{l,n}^{Z}\sigma_{l,s}^{Z}\sigma_{l,e}^{Z}\sigma_{l,w}^{Z}$,
in which the $(l,n)$, $(l,s)$, $(l,e)$, and $(l,w)$ label the physical spin
involved in the $l$th four-body constraint, and the interaction strength
should be the dominant energy scale in the embedding model \cite{R34,R36}. Therefore, for
a physical implementation of the LHZ scheme with four-body constraints,
the Ising problem Hamiltonian reads \cite{R34}
\begin{eqnarray}
\begin{aligned}
H_{P}^{LHZ_4}=\sum_{j=1}^{N_{P}}h_{j}\sigma_{j}^{Z}-C\sum_{l=1}^{N_{C}}\sigma_{l,n}^{Z}\sigma_{l,s}^{Z}\sigma_{l,e}^{Z}\sigma_{l,w}^{Z}
\end{aligned}
\end{eqnarray}
with logical field $h_{j}$ acted on the $j$th physical spin and $C$ the magnitude
of the four-spin coupling strength.

However, the four-spin interaction is still hard to be physically realized with
considerably high coupling strength. Recently, Leib \textit{et al.}$\,$ \cite{R35} have
theoretically demonstrated that one can use a general recursive decomposition
of classical k-local Ising terms, decomposing the four-body constraints into
two three-body constraints,
\begin{eqnarray}
\begin{aligned}
\sigma_{l,n}^{Z}\sigma_{l,s}^{Z}\sigma_{l,e}^{Z}\sigma_{l,w}^{Z}\rightarrow
\sigma_{l,n}^{Z}\sigma_{l,w}^{Z}\sigma_{l,a}^{Z}+\sigma_{l,a}^{Z}\sigma_{l,s}^{Z}\sigma_{l,e}^{Z}.
\end{aligned}
\end{eqnarray}
with an ancillary physical spin labeled by $(l,a)$. It is noted here that each ancillary
physical spin does not encode any information of the logical spin configuration, but mediates the
three-body realization of the four-body constraints. Therefore, one can realize the LHZ scheme with three-body
constraints, which can be realized directly by using a three-body interaction
$\sigma_{i}^{Z}\sigma_{j}^{Z}\sigma_{k}^{Z}$. For a physical implementation of
the LHZ scheme with three-body constraints, the Ising problem Hamiltonian becomes
\begin{eqnarray}
\begin{aligned}
H_{P}^{LHZ_3}=\sum_{j=1}^{\tilde{N}_{P}}h_{j}\sigma_{j}^{Z}
-C\sum_{l=1}^{\tilde{N}_{C}}\sigma_{l,i}^{Z}\sigma_{l,j}^{Z}\sigma_{l,k}^{Z},
\end{aligned}
\end{eqnarray}
with $\tilde{N}_{P}=N(N-1)+1$, and $\tilde{N}_{C}=(N-1)^{2}$.

In order to evaluate the performance of the LHZ scheme with three-spin interactions,
we present in Fig.$\,$5 the time-dependent energy spectrum of the executing
Hamiltonian for the Ising problem with $N=3$ logical spins, and the Ising problem
Hamiltonian are represented by Eq.$\,$(14), Eq.$\,$(16), and Eq.$\,$(18), respectively.
In the numerical simulation, we consider that the initial Hamiltonian
is given as $H_{I}=\sum_{j=1}^{N}b_{j}\sigma_{j}^{Z}$ with $b_{j}=\mathcal{J}$, and we
use the local field strength $h_{j}$, spin-spin coupling strength $\mathcal{J}_{jk}$
that is randomly taking from the interval $[-\mathcal{J},\mathcal{J}]$, and the constraints strength
$C/\mathcal{J}=3$. As shown in Fig.$\,$5, although different trajectories of
the time-dependent spectrum have been shown for the three schemes, an almost perfect
agreement of the lower $2^{3}$ energy levels is displayed at time $t=T$, demonstrating
preliminarily the validity of the three-body constraints implementation of the
Ising problem.

In the practical quantum annealing process, the minimal gap is the leading limitation
restricting the sweep time and also the main source of errors in quantum annealing,
i.e., Landau-Zener transitions. For the LHZ scheme with three-body- and four-body constraints,
we further presents in Fig.$\,$6(a) the comparison of the minimal gap. For $500$ random instances,
the three-body constraints based implementation does not considerably decrease the minimal gap,
and even the two implementations show a similar result for four different constraint
strengths ($C/\mathcal{J}=1.5,2,2.5,3$).

Alternatively, compared with the ramp protocol described by the Hamiltonian Eq.$\,$(15) where
in the annealing process, the constraint terms are adiabatically changed form $0$ to $C$,
it is also possible to use an always-on protocol (i.e., the strength of the constraints are
fixed in the whole process). Using this always-on protocol, the Ising problem encoded
by the LHZ scheme is fully characterized by the local fields applied on the physical spin,
and thus one can simplify the physical system and reduce the complexity for
quantum annealing. Recent theoretical studies have demonstrated that the gap in the
always-on protocol is in general smaller, but with only a small systematic difference
between the two in favor of the ramp protocol \cite{R35}. However, as shown in Fig.$\,$6(b),
the minimal gap in the three-body implementation with ramp protocol $[\Delta_{min}^{3}]_{ramp}$ is
general larger than the one in the four-body always-on protocol $[\Delta_{min}^{4}]_{on}$.
This suggests that the three-body implementation with ramp protocol may have
better performance than the four-body always-on protocol. In the following discussion,
we propose a resonator based system for implementing the LHZ scheme with three-body
constraints, and the tunable three-body interactions in the proposed system allows
for ramp protocol.

\begin{figure}[tbp]
\begin{center}
\includegraphics[width=8cm,height=6.0cm]{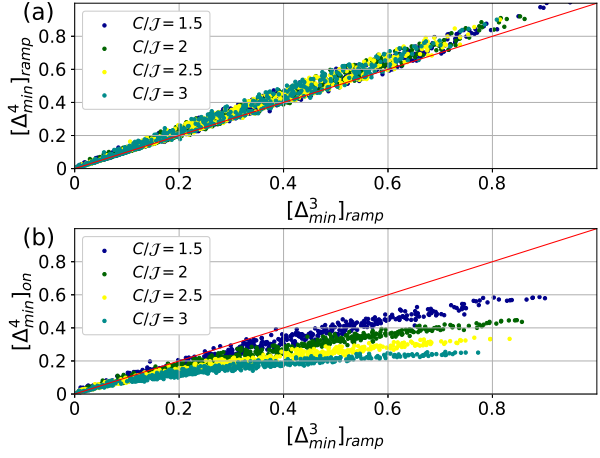}
\end{center}
\caption{(Color online) Scatter plot showing the minimal gap in the annealing process
for the LHZ scheme with three-spin interaction $\Delta_{min}^{3}$ versus the minimal
gap with four-spin interaction $\Delta_{min}^{4}$ .
The points (belong to the same color) correspond to 500 random instances with a given
constraints strength. The constraints strength are $C/\mathcal{J}=(1.5,2,2.5,3)$, and
the parameters of the Ising problem used are the same as that of Fig.$\,$5.
(a) For both LHZ implementations,
we use ramp protocol, i.e., $[\Delta_{min}^{3}]_{ramp}$ versus $[\Delta_{min}^{4}]_{ramp}$.
(b) The ramp protocol for the three-spin interaction implementation $[\Delta_{min}^{3}]_{ramp}$
against the always-on protocol for the four-spin interaction implementation
$[\Delta_{min}^{4}]_{on}$. }
\end{figure}

\subsection{The coupled network with application to the Lechner-Hauke-Zoller annealer}
\begin{figure}[tbp]
\begin{center}
\includegraphics[width=8cm,height=12.0cm]{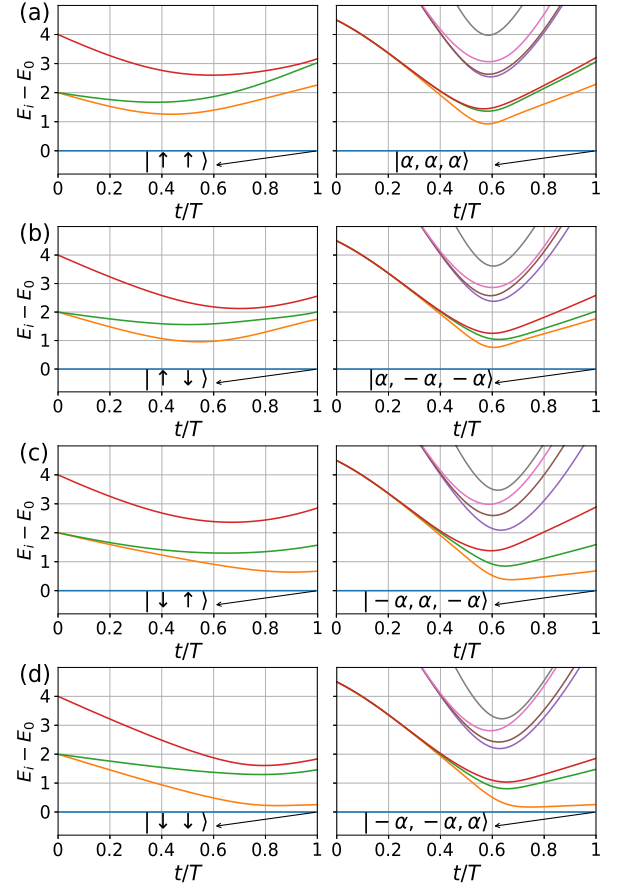}
\end{center}
\caption{(Color online) Time-dependent spectrum. Left panel: Energy
spectrum during a adiabatic evolution for Ising problem with $N=2$
logical spins, represented by shallow green block in Fig.$\,$4(a).
Right panel: Resonator based LHZ-scheme with three-body constraints
described by Eq.$\,$(26), represented by shallow green block in Fig.$\,$4(b). The parameters
of the Ising problem used are the same as that of Fig.$\,$5, and we use the constraints strength $C/\mathcal{J}=3$.
Resonator system parameters for calculation are $\delta_{1}/\mathcal{J}=\delta_{2}/\mathcal{J}=\delta_{3}/\mathcal{J}=4.5$,
$K/\mathcal{J}=10$, $\varepsilon_{j}^{(p)}/\mathcal{J}=-20$, resulting
$\alpha=\sqrt{P/K}=\sqrt{2}$, and $J_{123}/\mathcal{J}\approx-0.53$. In left panel of (a)-(d),
the arrowed line marks the spin configuration of the ground state at time $t=T$, i.e., $\{|\uparrow\uparrow\rangle,|\uparrow\downarrow\rangle,|\downarrow\uparrow\rangle,|\downarrow\downarrow\rangle\}$,
respectively, and also in the right panel, where the state of the first two resonators
reproduces the spin configuration of the ground state of the Ising problem, while the
state of the third resonator confirms the relative orientation of the two logical spins.}
\end{figure}

Following the recent theoretical studies in Ref.$\,$[21], here we show that the
cQED architecture proposed in Sec.II, can be used to realize
the LHZ scheme with three-body constraints, where the spin states $\{|\uparrow\rangle,|\downarrow\rangle\}$ are
encoded in the degenerate ground subspace of the two-photon driven Kerr resonator
formed by two coherent states of opposite phases, i.e, $\{|\alpha\rangle,|-\alpha\rangle\}$, respectively, and
the three-body constraints are directly implemented with three-resonator
interactions mediated by the transmon qubit, as we introduced in Sec.$\,$II(B).
Moreover, the proposal allows tunable three-resonator interactions that enables
the ramp protocol for the LHZ annealer.

For the triangular network as shown in Fig.$\,$4(b), we consider that each
spin is physically realized by a two-photon driven Kerr Resonator,
and all the four-body constraints are physically realized by two three-resonator
interactions with an ancillary resonator, as shown in Fig.$\,$4(d), while the
three-body constraints in the base of the lattice is directly realized by
three-resonator interactions. Meanwhile, the local field acted on the physical
spin is introduced by a single-photon drive applied on the resonator \cite{R20,R21}.
To implement the adiabatic ramp protocol for a general $N$-spin Ising problem with
the proposed triangular network, the executing time-dependent Hamiltonian
reads (in the lab frame)
\begin{eqnarray}
\begin{aligned}
H^{LHZ}_{N}(t)=&\sum_{j=1}^{\tilde{N}_{P}}(H_{j}+H_{j}^{(d)})+\sum_{(i,j,k)\in l}^{\tilde{N}_{C}}H_{ijk}^{C}
\end{aligned}
\end{eqnarray}
where
\begin{eqnarray}
\begin{aligned}
H_{j}=&\omega_{j}a^{\dag}_{j}a_{j}+K_{j}a^{\dag2}_{j}a_{j}^{2}
\\&-\varepsilon_{j}^{(p)}(t)(a^{\dag2}_{j}e^{-i\omega_{j}^{(p)}(t)t}+a_{j}^{2}e^{i\omega_{j}^{(p)}(t)t}),
\end{aligned}
\end{eqnarray}
describes the $j$th resonator with self-Kerr coefficient $K_{j}$ and a two-photon
drive of amplitude $\varepsilon_{j}^{(p)}(t)=\varepsilon_{j}^{(p)}t/T$ at frequency $\omega_{j}^{(p)}(t)=2\omega_{j}-2\delta_{j}(1-\frac{t}{2T})$, similar to the one
given in Sec.$\,$II(A),
\begin{eqnarray}
\begin{aligned}
H_{j}^{(d)}=\varepsilon_{j}^{(d)}(t)(a^{\dag}_{j}e^{-i\omega_{j}^{(d)}(t)t}+a_{j}e^{i\omega_{j}^{(d)}(t)t})
\end{aligned}
\end{eqnarray}
represents the additional single-photon drive of amplitude
$\varepsilon_{j}^{(d)}(t)=\varepsilon_{j}^{(d)}t/T$ applied on the $j$th resonator
at frequency $\omega_{j}^{(d)}(t)=\omega_{j}^{(p)}(t)/2$, and
\begin{eqnarray}
\begin{aligned}
H_{ijk}^{C}=J_{ijk}(t)(a^{\dag}_{i}a^{\dag}_{j}a_{k}e^{-i\omega_{d}^{(l)}t}+a_{i}a_{j}a_{k}^{\dag}e^{i\omega_{d}^{(l)}t})
\end{aligned}
\end{eqnarray}
denotes the three-resonator interaction with strength $J_{ijk}(t)=J_{ijk}t/T$,
which is induced by a pump mode at frequency $\omega_{d}^{(l)}=[\omega_{i}^{(d)}(t)+\omega_{j}^{(d)}(t)-\omega_{k}^{(d)}(t)]_{(i,j,k)\in l}$,
as demonstrated in Sec.II(B).

In a frame where each of the resonator rotate at the instantaneous single-photon
drive frequency, i.e., applying the unitary transformation
$U=e^{-i\sum_{j=1}^{\tilde{N}_{P}}\omega_{j}^{(d)}(t)ta^{\dag}_{j}a_{j}}$ on above expression,
the Hamiltonian now reads \cite{R21}
\begin{eqnarray}
\begin{aligned}
H^{LHZ}_{N}(t)=(1-\frac{t}{T})H_{I}+(\frac{t}{T})H_{P}^{LHZ_{R}}
\end{aligned}
\end{eqnarray}
where
\begin{eqnarray}
\begin{aligned}
H_{I}=\sum_{j=1}^{\tilde{N}_{P}}[(\delta_{j}a^{\dag}_{j}a_{j}+K_{j}a^{\dag2}_{j}a_{j}^{2})
\end{aligned}
\end{eqnarray}
acts as the initial Hamiltonian for implementing quantum annealing, whose ground state
is vacuum state that is actually simpler to be prepared, and
\begin{eqnarray}
\begin{aligned}
H_{P}^{LHZ_{R}}=&\sum_{j=1}^{\tilde{N}_{P}}[K_{j}a^{\dag2}_{j}a_{j}^{2}+
\varepsilon_{j}^{(p)}(a^{\dag2}_{j}+a_{j}^{2})
+\varepsilon_{j}^{(d)}(a^{\dag}_{j}+a_{j})]
\\&+\sum_{(i,j,k)\in l}^{\tilde{N}_{C}}J_{ijk}(a^{\dag}_{i}a^{\dag}_{j}a_{k}+a_{i}a_{j}a_{k}^{\dag})
\end{aligned}
\end{eqnarray}
characterizes the Ising problem by using the LHZ scheme with three-body constraints.
The correspondence between the above expression and the Hamiltonian given in Eq.$\,$(18),
can be found directly by projecting Eq.$\,$(25) on the tensor product
(i.e., $\bigotimes\tilde{N}_{P}$) of the spaces spanned by $\{|\alpha\rangle,|-\alpha\rangle\}$,
where the $\alpha$ is real number, and by dropping constant terms, resulting
\begin{eqnarray}
\begin{aligned}
H_{P}^{LHZ_{R}}=\sum_{j=1}^{\tilde{N}_{P}}h_{j}\sigma_{j}^{Z}
+\sum_{(i,j,k)\in l}^{\tilde{N}_{C}}C_{ijk}\sigma_{l,i}^{Z}\sigma_{l,j}^{Z}\sigma_{l,k}^{Z}
\end{aligned}
\end{eqnarray}
with $h_{j}=2\varepsilon_{j}^{(d)}\alpha$ and $C_{ijk}=2J_{ijk}\alpha^{3}$. However, as mentioned
in Ref.$\,$[52] and demonstrated in Refs.$\,$[13, 21], the above procedure is valid only if the
single-photon driven strength $\varepsilon_{j}^{d}$ is sufficiently smaller,
and also the three-resonator coupling strength $J_{ijk}$.

To give a preliminary verification of the above resonator based LHZ annealer,
we present in Fig.$\,$7 the time-dependent energy spectrum of the Hamiltonian in Eq.$\,$(15)
and Eq.$\,$(25), respectively, with $N=2$. At the end of the evolution $t=T$, an almost
perfect agreement of the lower $2^{2}$ energy levels is displayed. Furthermore,
in the left panel of Fig.$\,$(7), we present the the spin configuration of the
ground state of Ising problem described by Eq.$\,$(15) with different parameters,
resulting
$\{|\uparrow\uparrow\rangle,|\uparrow\downarrow\rangle,|\downarrow\uparrow\rangle,|\downarrow\downarrow\rangle\}$
for (a)-(d), respectively. We also show, in the right panel, the ground state of
the corresponding resonator-based LHZ annealer at time $t=T$, where the state of the
first two resonators reproduces the spin configuration of the ground state
of the Ising problem, while the state of the third resonator confirms the relative
orientation of the two logical spins. The whole fidelity
$\mathcal{F}= \langle s_{1}\alpha,s_{2}\alpha,s_{3}\alpha|\Psi_{0}(T)\rangle$,
in which $\alpha=\sqrt{P/K}=\sqrt{2}$, $s_{i}=\pm\,(i=1,2,3)$ is the sign of the amplitude
of the $i$th resonator for an ideal encoding, and $|\Psi_{0}(T)\rangle$ is the
eigenstate of the executing Hamiltonian Eq.$\,$(23) at $t=T$, are $99.93\%,99.93\%,99.93\%,99.94\%$
for the four different spin configurations, respectively.

\section{conclusion}

In conclusion, we propose a new scheme for implementing two-photon driven
Kerr resonator in a cQED architecture consisting of a superconducitng
resonator capacitively coupled to a microwave driven superconducting qubit,
and by using realistic parameters, we show that the Schr\"{o}dinger cat state
can be prepared via adiabatic evolution in the microwave-driven qubit-resonator system.
Contrary to the implementation by using superconducting coplanar resonator
terminated by a flux-pumped SQUID, our protocol is compatible with the 3D
architecture that allows us to exploits the large coherence times of the
3D microwave superconducting resonator. The major limitation of our proposal
is the population leakage for lager drive amplitudes, which results
infidelity, and in principle, this can be further decreased by looking for
optimal parameter set of the qubit-resonator system. As a possible extension,
one may also apply this protocol to a hybrid quantum system consisting of a
spin or atomic ensemble coupled to a superconducting circuits \cite{R53}, allowing
generation of spin (atomic) Schr\"{o}dinger cat state and spin squeezing for
quantum enhanced sensing and metrology \cite{R54,R55}.

Inspired by the recent theoretical work, we further show that the presented architecture
to realize two-photon driven Kerr resonator together with the introduced tunable
three-resonator interaction, can be scaled easily up to a coupled network for the implementation
of resonator based LHZ annealer. The tunable three-resonator interaction
allows for the ramp protocol of our quantum annealer. Therefore, comparing with the
always-on protocol, our implementation with ramp protocol
may have better performance. Furthermore, in principle, the annealer that we have proposed can be realized in
the 3D cQED architecture \cite{R56,R57}, for which the coherence times of the 3D microwave
superconducting resoantor can exceed that of the best superconducting qubit by almost two
orders of magnitude \cite{R31,R32}. This makes our proposed implementation a promising physical platform
for realizing quantum annealer with improved coherence.

\acknowledgements We thank Chui-Ping Yang for useful discussions. Thanks also to
the developers of the PYTHON package QuTiP \cite{R58,R59}, which was used for numerical calculation.
This work was partly supported by the NKRDP of China
(Grant No. 2016YFA0301802) and NSFC (Grants No. 11504165, No. 11474152, and No.
61521001).
\appendix

\section{The realistic circuit-QED implementation }

\begin{figure}[tbp]
\begin{center}
\includegraphics[width=8cm,height=3cm]{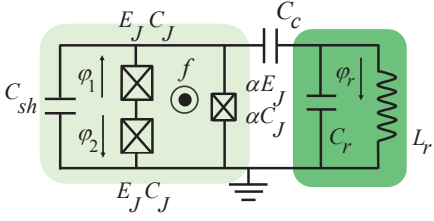}
\end{center}
\caption{(Color online) Circuit model of the capacitively-shunted flux qubit
(C-shunt flux qubit) capacitively coupled to an LC resonator.}
\end{figure}

Figure 8 shows the circuit model of the capacitively-shunted flux qubit
(C-shunt flux qubit) \cite{R60,R61}, which is capacitively coupled to an LC resonator \cite{R61,R62,R63}.
The two identical Josephson junctions have capacitance $C_{J}$ and
coupling energy $E_{J}$, while the third (smaller one) has
capacitance $\alpha C_{J}$ and coupling energy $\alpha E_{J}$.
$L_{r}$ and $C_{r}$ represent the equivalent inductance and
capacitance of the resonator, respectively. $C_{c}$ represents
the coupling capacitance between the resonator and the flux qubit,
for which the smaller junction is shunted by a capacitance $C_{sh}$.
In the following discussion, the phase difference across the larger
Josephson junctions and the inductance $L_{r}$ are denoted as
$\varphi_{i}$ $(i=1, 2)$ and $\varphi_{r}$, respectively. The
Lagrangian of the system is given by
\begin{eqnarray}
\begin{aligned}
\mathcal{L}=&\frac{C_{J}}{2}(\frac{\Phi_{0}}{2\pi})^{2}\dot{\varphi}_{1}^{2}+
\frac{C_{J}}{2}(\frac{\Phi_{0}}{2\pi})^{2}\dot{\varphi}_{2}^{2}
\\&+\frac{\alpha C_{J}}{2}(\frac{\Phi_{0}}{2\pi})^{2}(\dot{\varphi}_{1}^{2}-\dot{\varphi}_{2}^{2})
\\&+\frac{C_{sh}}{2}(\frac{\Phi_{0}}{2\pi})^{2}(\dot{\varphi}_{1}^{2}-\dot{\varphi}_{2}^{2})+U
\\&+\frac{C_{c}}{2}(\frac{\Phi_{0}}{2\pi})^{2}(\dot{\varphi}_{r}^{2}+\dot{\varphi}_{1}^{2}-\dot{\varphi}_{2}^{2})
\\&+\frac{C_{r}}{2}(\frac{\Phi_{0}}{2\pi})^{2}\dot{\varphi}_{r}^{2}+\frac{1}{2L_{r}}(\frac{\Phi_{0}}{2\pi})^{2}\varphi_{r}^{2}
\end{aligned}
\end{eqnarray}
where $\Phi_{0}=h/2e$, and $U=E_{J}\cos\varphi_{1}+E_{J}\cos\varphi_{2}
+\alpha E_{J}\cos(\varphi_{1}-\varphi_{2}+2\pi f)$ with $f=\Phi_{e}/\Phi_{0}$,
where $\Phi_{e}$ is the externally applied magnetic flux in the loop.
For clarity, this equation can be rewritten as
\begin{eqnarray}
\begin{aligned}
\mathcal{L}=&\frac{C_{J}}{2}(\frac{\Phi_{0}}{2\pi})^{2}\dot{\varphi}_{1}^{2}+
\frac{C_{J}}{2}(\frac{\Phi_{0}}{2\pi})^{2}\dot{\varphi}_{2}^{2}
\\&+\frac{\alpha^{\prime} C_{J}}{2}(\frac{\Phi_{0}}{2\pi})^{2}(\dot{\varphi}_{1}^{2}-\dot{\varphi}_{2}^{2})
\\&+U+\frac{C_{c}}{2}(\frac{\Phi_{0}}{2\pi})^{2}(\dot{\varphi}_{r}^{2}+\dot{\varphi}_{1}^{2}-\dot{\varphi}_{2}^{2})
\\&+\frac{C_{r}}{2}(\frac{\Phi_{0}}{2\pi})^{2}\dot{\varphi}_{r}^{2}+\frac{1}{2L_{r}}(\frac{\Phi_{0}}{2\pi})^{2}\varphi_{r}^{2}
\end{aligned}
\end{eqnarray}
with $\alpha^{\prime}=\alpha+\frac{C_{sh}}{C_{J}}$. Following the procedure of Ref.$\,$[63], the
system can described by a Hamiltonian consisting of three parts,
\begin{eqnarray}
\begin{aligned}
H_{q_r}=H_{q}+H_{r}+H_{c},
\end{aligned}
\end{eqnarray}
where
\begin{eqnarray}
\begin{aligned}
H_{q}=&4E_{c}\frac{(1+\alpha^{\prime})(1+\gamma)+\beta}{(1+2\alpha^{\prime})(1+\gamma)+2\beta}(n_{1}^{2}+n_{2}^{2})
\\&+8E_{c}\frac{\alpha^{\prime}(1+\gamma)+\beta}{(1+2\alpha^{\prime})(1+\gamma)+2\beta}n_{1}n_{2}-U,
\end{aligned}
\end{eqnarray}
is the qubit Hamiltonian,
\begin{eqnarray}
\begin{aligned}
H_{r}=\omega _{c}a^{\dagger}a
\end{aligned}
\end{eqnarray}
is the resonator Hamiltonian, and
\begin{eqnarray}
\begin{aligned}
H_{c}=&\frac{-2i}{(1+2\alpha^{\prime}+2\beta)^{1/4}}\sqrt{\frac{\beta\gamma}{[(1+2\alpha^{\prime})(1+\gamma)+2\beta]^{3/2}}}
\\&\times \sqrt{E_{r}E_{c}}(n_{1}-n_{2})(a^{\dagger}-a)
\end{aligned}
\end{eqnarray}
is the qubit-resonator interaction Hamiltonian. Above, $a^{\dagger }$ and $a$
are the creation and annihilation operators for the resonator with frequency
\begin{eqnarray}
\begin{aligned}
\omega_{c}=\frac{1}{\sqrt{L_{r}C_{r}}}\sqrt{\frac{1+2\alpha^{\prime}
+2\beta}{(1+2\alpha^{\prime})(1+\gamma)+2\beta}},
\end{aligned}
\end{eqnarray}
and $E_{c}=e^{2}/2C_{J}$, $\beta=C_{c}/C_{J}$, $\gamma=C_{c}/C_{r}$, and $n_{i}$ $(i=1, 2)$ are conjugate
variable of the phase differences $\varphi_{i}$.

\begin{table}[htbp]
\caption{Component parameters of the qubit-resonator system used for energy level and coupling strength calculations in Fig.$\,$9.}
\begin{tabular}{p{2.6cm}<{\centering}p{4.0cm}<{\centering}}
\hline
\hline
$C_{J} $& 10.76 fF\\
$E_{J}/2\pi $& 135.00 GHz\\
$\alpha $& 0.60 \\
$C_{sh} $& 22.06 fF\\
$C_{c}$& 5.92 fF\\
$E_{r}/2\pi $& 5.25 GHz \\
\hline
\hline
\end{tabular}
\end{table}
\begin{figure}[tbp]
\begin{center}
\includegraphics[width=8cm,height=6cm]{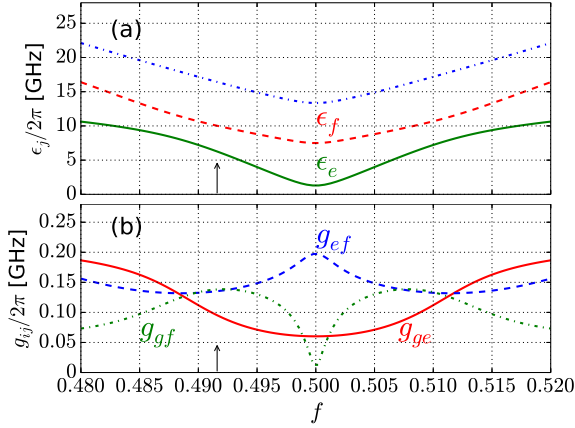}
\end{center}
\caption{(Color online) (a) The energy spectrum of the flux qubit from the ground state.
(b) The coupling strength $g_{ij}=|\langle i|H_{c}|j\rangle|$. In this work, the biased
point of interest to us is indicated by the vertical arrow, i.e., $f=0.4916$. This
results the system parameters used for numerical analysis in the main text,
$\varepsilon_{e}/2\pi=6.25$ GHz, $\varepsilon_{f}/2\pi=10.0$ GHz,
$g_{ge}/2\pi=0.094$ GHz, $g_{ef}/2\pi=0.136$ GHz, and $g_{gf}/2\pi=0.140$ GHz.}
\end{figure}

According to the parameters listed in Table 1, Figure 9 shows the
energy spectrum of the flux qubit from the ground state and
the coupling strength $g_{ij}=|\langle i|H_{c}|j\rangle|$, where
$|j\rangle$ represents the $j$th eigenstate of the uncoupled qubit
Hamiltonian given in Eq.$\,$(A4), as a function of the flux bias $f$.

In this work, we focus on the lowest three
energy levels of the flux qubit and thus in this situation the flux qubit
is considered to be a perfect three-level system (qutrit). After using the RWA,
the Hamiltonian of the qutrit-resonator system is
\begin{eqnarray}
\begin{aligned}
&H_{q_r}=\omega_{c}a^{\dagger}a+\sum_{j=g,e,f}\epsilon_{j}|j\rangle\langle j|+H_{c},
\\&H_{c}=g_{ge}|g\rangle\langle e|a^{\dagger}+g_{ef}|e\rangle\langle f|a^{\dagger}
+g_{gf}|g\rangle\langle f|a^{\dagger}+H.c.
\end{aligned}
\end{eqnarray}
By introducing a microwave drive tone applied on the qutrit, the full system Hamiltonian reads
\begin{eqnarray}
\begin{aligned}
H_{full}=H_{q_r}+H_{d},
\end{aligned}
\end{eqnarray}
where
\begin{eqnarray}
\begin{aligned}
H_{d}=&\Omega^{(p)}_{ge}e^{-i\omega_{p}t}|e\rangle\langle g|+
\Omega^{(p)}_{ef}e^{-i\omega_{p}t}|f\rangle\langle e|
\\&+\Omega^{(p)}_{gf}e^{-i\omega_{p}t}|f\rangle\langle g|+H.c.,
\end{aligned}
\end{eqnarray}
here $H.c.$ stands for Hermitian conjugate, and $\epsilon_{j}\,(j=g,e,f)$
is the transition frequency of the qutrit from ground to excited
state $|j\rangle$. $g_{ge}$, $g_{ef}$, and $g_{gf}$ denote
the qutrit-resonator coupling strength. $\Omega^{(p)}_{jk}$ is the real
amplitude of the microwave drive with frequency $\omega _{p}$ applied
to the $|j\rangle\longleftrightarrow|k\rangle$ transition of the qutrit.
For simplicity, we define $\epsilon_{g}=0$ in the following discussion.

We consider that our system operates in the dispersive regime, where
the qutrit is detuned from the resonator $|\Delta _{jk}|=|(|\epsilon_{j}
-\epsilon_{k}|)-\omega _{c}|\gg g_{jk}$, and the qutrit is
off-resonantly driven by the microwave drive $|\Delta _{jk}^{\prime }|
=|(|\epsilon_{j}-\epsilon_{k}|)-\omega_{p}|\gg \Omega^{(p)}_{jk}$. Moreover,
we suppose that the qutrit is initially in its ground state. Therefore,
the qutrit remains unexcited in the process and thus decouples from the
resonator. In these conditions, by using the fourth-order perturbation
theory \cite{R40,R64}, and eliminating the degrees of freedom of
the qutrit, the Hamiltonian in Eq.$\,$(A9) can be well approximated by the
effective hamiltonian
\begin{eqnarray}
\begin{aligned}
H_{eff}=\widetilde{\omega}_{c} a^{\dagger}a+Ka^{\dagger2}a^{2}-P(a^{\dagger 2}
e^{-i\omega_{p}t}+H.c.)
\end{aligned}
\end{eqnarray}
where $\widetilde{\omega}_{c}=\omega_{c}+S$ is dressed resonator
frequency with
\begin{eqnarray}
\begin{aligned}
S=&-\frac{g_{ge}^{2}}{\omega_{e}-\omega_{c}}-\frac{g_{gf}^{2}}{\omega_{f}-\omega_{c}}
+\frac{g_{ge}^{4}}{(\omega_{e}-\omega_{c})^{3}}\\&+\frac{g_{gf}^{4}}{(\omega_{f}-\omega_{c})^{3}}
+\frac{g_{gf}^{2}g_{ef}^{2}}{(\omega_{c}-\omega_{f})(-\omega_{e})(\omega_{c}-\omega_{f})}
\\&-\frac{g_{ge}^{2}g_{gf}^{2}}{(\omega_{c}-\omega_{e})^{2}(\omega_{c}-\omega_{f})}
-\frac{g_{ge}^{2}g_{gf}^{2}}{(\omega_{c}-\omega_{e})(\omega_{c}-\omega_{f})^{2}}
\end{aligned}
\end{eqnarray}
The second term denotes the qutrit-induced Kerr-nonlinearity of the
resonator with
\begin{eqnarray}
\begin{aligned}
K=&\frac{g_{ge}^{4}}{(\omega_{e}-\omega_{c})^{3}}+\frac{g_{gf}^{4}}{(\omega_{f}-\omega_{c})^{3}}
\\&+\frac{g_{ge}^{2}g_{ef}^{2}}{(\omega_{c}-\omega_{e})(2\omega_{c}-\omega_{f})(\omega_{c}-\omega_{e})}
\\&+\frac{g_{gf}^{2}g_{ef}^{2}}{(\omega_{c}-\omega_{f})(-\omega_{e})(\omega_{c}-\omega_{f})}
\\&-\frac{g_{ge}^{2}g_{gf}^{2}}{(\omega_{c}-\omega_{e})^{2}(\omega_{c}-\omega_{f})}
-\frac{g_{ge}^{2}g_{gf}^{2}}{(\omega_{c}-\omega_{e})(\omega_{c}-\omega_{f})^{2}}
\end{aligned}
\end{eqnarray}
The last term represents a two-photon drive of amplitude
\begin{eqnarray}
\begin{aligned}
P=-\frac{g_{ge}g_{ef}\Omega^{(p)}_{gf}}{\Delta_{ge}(\epsilon_{f}-\omega_{p})}
\end{aligned}
\end{eqnarray}
applied on the resonator at frequency $\omega_{p}=2\widetilde{\omega}_{c}$.

\section{master equation}

The influence of photon decay and qubit relaxation on the quantum adiabatic evolution
can be studied by the master equation approach. By including photon decay and qubit
relaxation terms, we can write the master equation \cite{R65}
\begin{eqnarray}
\begin{aligned}
\frac{d\rho_{s}}{dt}=&-i[H,\rho_{s}]+\kappa\mathcal{L}[a]+\gamma_{ge}\mathcal{L}[|g\rangle _{q}\langle e|]\\&+\gamma_{ef}\mathcal{L}[|e\rangle _{q}\langle f|]+\gamma_{gf}\mathcal{L}[|g\rangle _{q}\langle f|].
\end{aligned}
\end{eqnarray}
Above, $\rho_{s}$ is the density matrix of the entire system, $H$ is the
Hamiltonian of the full system given in Eq.$\,$(A9), and $\mathcal{L}[O]=O\rho O^{\dagger }-O^{\dagger }
O\rho/2-\rho O^{\dagger }O/2$ is the Lindbladian of the operator $O$. $\kappa $ and $\gamma _{jk}$ denote
the photon decay rate of the resonator and the relaxation rate of the $(|j\rangle,|k\rangle)$
two level systems, respectively.

To make a direct comparison of the evolution of the effective model describing by
Hamiltonian in Eq.$\,$(9) in the main text and that of our proposed full microwave-driven qutrit-resonator
system describing by the Hamiltonian in Eq.$\,$(A9), it is helpful to move to the rotating frame
corresponds to the interaction picture with respect to the
renormalized frequency of the resonator, that is the change of variables in Eq.$\,$(B1)
\begin{eqnarray}
\begin{aligned}
\tilde{\rho_{s}}(t)=e^{i\tilde{\omega}_{c}t}\rho_{s} e^{-i\tilde{\omega}_{c}t}
\end{aligned}
\end{eqnarray}
In this new frame, the master equation is given as
\begin{eqnarray}
\begin{aligned}
\frac{d\tilde{\rho_{s}}}{dt}=&-i[\tilde{H},\tilde{\rho_{s}}]+
\kappa\tilde{\mathcal{L}}[\tilde{a}]+\gamma_{ge}\tilde{\mathcal{L}}[|g\rangle \langle e|]
\\&+\gamma_{ef}\tilde{\mathcal{L}}[|e\rangle\langle f|]+
\gamma_{gf}\tilde{\mathcal{L}}[|g\rangle\langle f|],
\end{aligned}
\end{eqnarray}
where $\tilde{\mathcal{L}}[O]=O\tilde{\rho_{s}} O^{\dagger }-O^{\dagger }
O\tilde{\rho_{s}}/2-\tilde{\rho_{s}} O^{\dagger }O/2$, and $\tilde{H}=\tilde{H}_{q_r}+H_{d}$ with
\begin{eqnarray}
\begin{aligned}
&\tilde{H}_{q_r}=(\omega_{c}-\tilde{\omega}_{c})a^{\dagger}a+\sum_{j=g,e,f}\epsilon_{j}|j\rangle\langle j|+\tilde{H}_{c}
\\&\tilde{H}_{c}=g_{ge}|g\rangle\langle e|\tilde{a}^{\dagger}+g_{ef}|e\rangle\langle f|\tilde{a}^{\dagger}
+g_{gf}|g\rangle\langle f|\tilde{a}^{\dagger}+H.c.
\end{aligned}
\end{eqnarray}
Here, $\tilde{a}^{\dagger}=e^{i\tilde{\omega}_{c}t}a^{\dagger}$ and $\tilde{a}=e^{-i\tilde{\omega}_{c}t}a$
are the creation and annihilation operators for the resonator in the rotating frame.

\section{three-body interaction among the three resonators}
Here, we present a detailed derivation of the effective Hamiltonian Eq.$\,$(13)
in the main text. In the displaced frame with respect to the unitary transformation
\begin{eqnarray}
\begin{aligned}
&U(t)=e^{-\tilde{\xi}_{p}a^{\dag}_{q}+\tilde{\xi}^{\ast}_{p}a_{q}},
\\&\tilde{\xi}_{p}=\xi_{p}e^{-i\omega_{d}t},\,\,\,\,\xi_{p}=\frac{\varepsilon_{p}}{\omega_{d}-\omega_{q}},
\end{aligned}
\end{eqnarray}
with $U(t)aU(t)^{-1}=a+\tilde{\xi}_{p}$, the Hamiltonian in Eq.$\,$(12) reads
\begin{eqnarray}
\begin{aligned}
&H=\omega^{(0)}_{q}a^{\dag}_{q}a_{q}+\sum_{j=1}^3\omega^{(0)}_{j}a^{\dag}_{j}a_{j}-E_{J}(\cos{\varphi}+\frac{1}{2}\varphi^{2}),
\\&\varphi=[\phi_{q}(\tilde{a}^{\dag}_{q}+\tilde{a}_{q}+\tilde{\xi}^{\ast}_{p}+\tilde{\xi}_{p})+\sum_{j=1}^3\phi_{j}(a^{\dag}_{j}+a_{j})],
\end{aligned}
\end{eqnarray}
where $\tilde{a}^{\dag}_{q}$ and $\tilde{a}_{q}$ are the creation and annihilation
operators for the qubit mode in the displaced frame. Assuming small phase
fluctuations, we can expand the cosine up to the fourth order and only keep
the nonrotating terms, leading to the effective Hamiltonian \cite{R30}
\begin{eqnarray}
\begin{aligned}
H=\omega'_{q}a^{\dag}_{q}a_{q}+\sum_{j=1}^3\omega'_{j}a^{\dag}_{j}a_{j}
-E_{J}(\frac{1}{24}\varphi^{4}+\mathcal{O}(\varphi^{6})),
\end{aligned}
\end{eqnarray}
\begin{eqnarray}
\begin{aligned}
H=&\omega'_{q}a^{\dag}_{q}a_{q}+K_{q}a^{\dag2}_{q}a_{q}^{2}+\sum_{j=1}^3K_{qj}a^{\dag}_{q}a_{q}a^{\dag}_{j}a_{j}
\\&+\sum_{j=1}^3(\omega'_{j}a^{\dag}_{j}a_{j}+K_{j}a^{\dag2}_{j}a_{j}^{2})+\sum_{j\neq k}K_{jk}a^{\dag}_{j}a^{\dag}_{k}a_{j}a_{k}
\\&+|\xi_{p}|^{2}(2K_{q}a^{\dag}_{q}a_{q}+\sum_{j=1}^3K_{qj}a^{\dag}_{j}a_{j})
\\&+J_{123}(a^{\dag}_{1}a^{\dag}_{2}a_{3}e^{-i\omega_{d}t}+a_{1}a_{2}a^{\dag}_{3}e^{i\omega_{d}t}),
\end{aligned}
\end{eqnarray}
where $\omega'_{m}$ ($m=q,1,2,3$) is the frequency for the $m$th mode including a
renormalization of the transition frequency coming from the normal ordering procedure
of the fourth-order phase term. $K_{q}=-E_{J}\phi_{q}^{4}/4$ and
$K_{j}=-E_{J}\phi_{j}^{4}/4$ are
the coefficients of the self-Kerr nonlinearity associated with the qubit mode and the $i$th
resonator, respectively. $K_{jk}=-E_{J}\phi_{j}^{2}\phi_{k}^{2}$
denotes the coefficient of the cross-kerr nonlinearity between the $j$th resonator
and the $k$th resonator, and $K_{qj}=-E_{J}\phi_{q}^{2}\phi_{j}^{2}$ represents
the coefficient of the cross-kerr nonlinearity between the qubit mode
and the $j$th resonator. $J_{123}=-E_{J}\phi_{1}\phi_{2}\phi_{3}\phi_{q}\xi_{p}$
is three-resonator coupling strength. The term in the third line of Eq.$\,$(C4)
corresponds to the AC Stark shift induced by the pump mode. Note that we have
not neglected the terms like $a^{\dag}_{1}a^{\dag}_{2}a_{3}$ as well as we will choose
$\omega_{d}=\omega_{1}+\omega_{2}-\omega_{3}$ so that these terms are resonant.

By assuming that the qubit-resonator system operates in strongly dispersive regime,
and the pump mode is far off-resonance tone applied on the qubit mode, the qubit mode,
which is initially in its ground state, will remain unexcited in the whole process.
Therefore, to simply the above expression, we can safely eliminate the degrees of the
freedom of the qubit mode, and rewrite the Hamiltonian as
\begin{eqnarray}
\begin{aligned}
H=&\sum_{j=1}^3(\omega_{j}a^{\dag}_{j}a_{j}+K_{j}a^{\dag2}_{j}a^{2}_{j})
+\sum_{j\neq k}K_{jk}a^{\dag}_{j}a^{\dag}_{k}a_{j}a_{k}
\\&+J_{123}(a^{\dag}_{1}a^{\dag}_{2}a_{3}e^{-i\omega_{d}t}+a_{1}a_{2}a^{\dag}_{3}e^{i\omega_{d}t}),
\end{aligned}
\end{eqnarray}
where $\omega_{j}=\omega'_{j}+|\xi_{p}|^{2}K_{qj}$.
For realistic system, the coefficient $K_{j}$ and $K_{jk}$ are very small,
and one can omit these associated terms in the Hamiltonian, thus we can recover the Hamiltonian
given in Eq.$\,$(13) of the main text. When the pump frequency matches the detuning of the
three resonators, i.e. $\omega_{d}=\omega_{1}+\omega_{2}-\omega_{3}$, in the interaction
picture, the Hamiltonian reads as
\begin{eqnarray}
\begin{aligned}
H_{3body}=J_{123}(a^{\dag}_{1}a^{\dag}_{2}a_{3}+a_{1}a_{2}a^{\dag}_{3}).
\end{aligned}
\end{eqnarray}
By taking $E_{J}/2\pi=21$ GHz, and
$(\phi_{q}, \phi_{1}, \phi_{2}, \phi_{3}, \xi_{p})=(0.35,0.03,0.03,0.03,0.5)$,
it is possible to obtain a three-resonator coupling strength $J_{123}/2\pi\approx-0.1$ MHz.
These parameters also yield $K_{j}/2\pi\approx-4.25$ KHz, and $K_{jk}/2\pi\approx-4.25$ KHz, which
is rather small, as excepted. Moreover, in our protocol for implementing the
resonator-based LHZ annealer, the qubit-induced self-Kerr nonlinearly is only a
small correction to the Kerr term given in Hamiltonian Eq.$\,$(19), while for the
cross-Kerr nonlinearly, it has been demonstrated that the error caused by this
nonlinearity terms is very small for larger $\alpha$, and can be compensated by
introducing additional detuning terms $\delta a^{\dag}_{j}a_{j}$ for the Ising
problem Hamiltonian \cite{R21}.


\begin{thebibliography}{99}

\bibitem{R1} B. Wielinga and G. J. Milburn, Quantum tunneling in a Kerr medium with parametric pumping,
Phys. Rev. A \textbf{48}, 2494 (1993).
\bibitem{R2} B. Wielinga and G. J. Milburn, Tunneling in the presence of driving in a cavity that contains
a Kerr medium and is parametrically pumped, Phys. Rev. A \textbf{49}, 5042 (1994).
\bibitem{R3} G. Y. Kryuchkyan and K. V. Kheruntsyan, Exact quantum theory of a parametrically driven
dissipative anharmonic oscillator, Opt. Commun. \textbf{127}, 230 (1996).
\bibitem{R4} W. Leo\'{n}ski, Fock states in a Kerr medium with parametric pumping, Phys. Rev. A \textbf{54},
3369 (1996).
\bibitem{R5} K. V. Kheruntsyan, D. S. Krahmer, G. Yu. Kryuchkyan, and K. G. Petrossian, Wigner function for
a generalized model of a parametric oscillator: phase-space tristability, competition and nonclassical effects,
Opt. Commun. \textbf{139}, 157 (1997).
\bibitem{R6} T. V. Gevorgyan and G. Yu. Kryuchkyan, Parametrically driven nonlinear oscillator at a few-photon
level, J. Mod. Opt. \textbf{60}, 860 (2013).
\bibitem{R7} C. H. Meaney, H. Nha, T. Duty, and G. J. Milburn, Quantum and classical nonlinear dynamics in a
microwave cavity, Eur. Phys. J. Quantum Technol. \textbf{1}, 7 (2014).
\bibitem{R8} G. H. Hovsepyan, A. R. Shahinyan, Lock Yue Chew, and G. Yu. Kryuchkyan, Phase locking and quantum
statistics in a parametrically driven nonlinear resonator, Phys. Rev. A \textbf{93}, 043856 (2016).
\bibitem{R9} N. Bartolo, F. Minganti, W. Casteels, and C. Ciuti, Exact steady state of a Kerr resonator with one-
and two-photon driving and dissipation: Controllable Wigner-function multimodality and dissipative phase transitions,
Phys. Rev. A \textbf{94}, 033841 (2016).
\bibitem{R10} F. Minganti, N. Bartolo, J. Lolli, W. Casteels, and C. Ciuti, Exact results for Schr\"{o}dinger cats
in driven-dissipative systems and their feedback control, Sci. Rep. \textbf{6}, 26987 (2016).
\bibitem{R11} X. L. Zhao, Z. C. Shi, M. Qin, and X. X. Yi, Optical Schr\"{o}dinger cat states in one mode and two
coupled modes subject to environments, Phys. Rev. A \textbf{96}, 013824 (2017).
\bibitem{R12} H. Goto, Bifurcation-based adiabatic quantum computation with a nonlinear oscillator network, Sci.
Rep. \textbf{6}, 21686 (2016).
\bibitem{R13} S. Puri, S. Boutin and A. Blais, Engineering the quantum states of light in a Kerr-nonlinear resonator
by two-photon driving, npj Quant. Informa. \textbf{3}, 18 (2017).
\bibitem{R14} N. Bartolo, F. Minganti, J. Lolli, and C. Ciuti, Homodyne versus photon-counting quantum trajectories
for dissipative Kerr resonators with two-photon driving, Eur. Phys. J. Spec. Top. \textbf{226} (12), 2705-2713 (2017).
\bibitem{R15} A. Gilchrist, K. Nemoto, W. J. Munro, T. C. Ralph, S. Glancy, S. L. Braunstein, and G. J. Milburn,
Schr\"{o}dinger cats and their power for quantum information processing, J. Opt. B: Quantum Semiclassical Opt. \textbf{6},
S828 (2004).
\bibitem{R16} M. Mirrahimi, Z. Leghtas, V. V. Albert, S. Touzard, R. J. Schoelkopf, L. Jiang, and M. H. Devoret,
Dynamically protected cat-qubits: a new paradigm for universal quantum computation, New J. Phys. \textbf{16},
045014 (2014).
\bibitem{R17} A. Das and B. K. Chakrabarti, Colloquium: Quantum Annealing and Analog Quantum Computation,
    Rev. Mod. Phys. \textbf{80}, 1061 (2008).
\bibitem{R18} F. Barahona, On the computational complexity of Ising spin glass models, J. Phys. A \textbf{15},
3241 (1982).
\bibitem{R19} A. Lucas, Ising Formulations of Many NP Problems, Front. Phys. \textbf{2}, 5 (2014).
\bibitem{R20} H. Goto , Z. R. Lin, Y. Nakamura, Dissipative quantum bifurcation machine: Quantum heating of coupled
nonlinear oscillators, arXiv:1707.00986 (2017).
\bibitem{R21} S. Puri, C. K. Andersen, A. L. Grimsmo, and A. Blais, Quantum annealing with all-to-all connected
nonlinear oscillators, Nat. Commun. \textbf{8}, 15785 (2017).
\bibitem{R22} S. E. Nigg, N. L\"{o}rch, and R. P. Tiwari, Robust quantum optimizer with full connectivity, Sci.
Adv. \textbf{3}, e1602273 (2017).
\bibitem{R23} G. Kirchmair, B. Vlastakis, Z. Leghtas, S. E. Nigg, H. Paik, E. Ginossar, M. Mirrahimi, L. Frunzio,
S. M. Girvin, and R. J. Schoelkopf, Observation of quantum state collapse and revival due to the single-photon Kerr
effect, Nature (London) \textbf{495}, 205 (2013).
\bibitem{R24} J. Bourassa, F. Beaudoin, J. M. Gambetta, and A. Blais, Josephson-junction-embedded transmission-line
resonators: From Kerr medium to in-line transmon, Phys. Rev. A \textbf{86}, 013814 (2012).
\bibitem{R25} S. E. Nigg, H. Paik, B. Vlastakis, G. Kirchmair, S. Shankar, L. Frunzio, M. H. Devoret, R. J. Schoelkopf,
and S. M. Girvin, Black-Box Superconducting Circuit Quantization, Phys. Rev. Lett. \textbf{108}, 240502 (2012).
\bibitem{R26} Y. Makhlin, G. Sch\"{o}n, and A. Shnirman, Quantum-state engineering with Josephson-junction devices,
Rev. Mod. Phys. \textbf{73}, 357 (2001).
\bibitem{R27} T. Yamamoto, K. Inomata, M. Watanabe, K. Matsuba, T. Miyazaki, W. D. Oliver, Y. Nakamura, and J. S. Tsai,
Flux-driven Josephson parametric amplifier, Appl. Phys. Lett. \textbf{93}, 042510 (2008).
\bibitem{R28} C. M. Wilson, T. Duty, M. Sandberg, F. Persson, V. Shumeiko, and P. Delsing, Photon Generation in an
Electromagnetic Cavity with a Time-Dependent Boundary, Phys. Rev. Lett. \textbf{105}, 233907 (2010).
\bibitem{R29} W. Wustmann and V. Shumeiko, Parametric resonance in tunable superconducting cavities, Phys. Rev.
B \textbf{87}, 184501 (2013).
\bibitem{R30} Z. Leghtas, S. Touzard, I. M. Pop, A. Kou, B. Vlastakis, A. Petrenko, K. M. Sliwa, A. Narla, S. Shankar,
M. J. Hatridge, M. Reagor, L. Frunzio, R. J. Schoelkopf, M. Mirrahimi, and M. H. Devoret, Confining the state of light
to a quantum manifold by engineered two-photon loss, Science \textbf{347}, 853 (2014).
\bibitem{R31} M. Reagor, H. Paik, G. Catelani, L. Sun, C. Axline, E. Holland, I. M. Pop, N. A. Masluk, T. Brecht, L.
Frunzio, M. H. Devoret, L. Glazman, and R. J. Schoelkopf, Reaching 10 ms single photon lifetimes for superconducting
aluminum cavities, Appl. Phys. Lett. \textbf{102}, 192604 (2013).
\bibitem{R32} M. Reagor, W. Pfaff, C. Axline, R. W. Heeres, N. Ofek, K. Sliwa, E. Holland, C. Wang, J. Blumoff, K. Chou,
M. J. Hatridge, L. Frunzio, M. H. Devoret, L. Jiang, and R. J. Schoelkopf, Quantum memory with millisecond coherence in
circuit QED, Phys. Rev. B \textbf{94}, 014506 (2016).
\bibitem{R33} M. Mirrahimi, Cat-qubits for quantum computation, Comptes Rendus Phys. \textbf{17}, 778 (2016).
\bibitem{R34} W. Lechner, P. Hauke, P. Zoller, A quantum annealing architecture with all-to-all connectivity from local
interactions. Sci. Adv. \textbf{1} , e1500838 (2015).
\bibitem{R35} M. Leib, P. Zoller, and W. Lechner, A transmon quantum annealer: Decomposing many-body
ising constraints into pair interactions, Quantum Science and Technology \textbf{1}, 015008 (2016).
\bibitem{R36} A. Rocchetto, S. C. Benjamin, and Y. Li, Stabilizers as a design tool for new forms of the
Lechner-Hauke-Zoller annealer. Sci. Adv. \textbf{2}, e1601246 (2016).
\bibitem{R37} T. Orlando, J. Mooij, L. Tian, C. van der Wal, L. Levitov, S. Lloyd, and J. Mazo, Superconducting
persistent-current qubit, Phys. Rev. B \textbf{60}, 15398 (1999).
\bibitem{R38} Y.-X. Liu, J. Q. You, L. F. Wei, C. P. Sun, and F. Nori, Optical selection rules and phase-dependent
adiabatic state control in a superconducting quantum circuit, Phys. Rev. Lett. \textbf{95}, 087001 (2005).
\bibitem{R39} V. E. Manucharyan, J. Koch, L. I. Glazman, and M. H. Devoret, Fluxonium: Single cooper-pair circuit free
of charge offsets, Science \textbf{326}, 113 (2009).
\bibitem{R40} G. Zhu, D. G. Ferguson, V. E. Manucharyan, and J. Koch, Circuit QED with fluxonium qubits: Theory of the
dispersive regime, Phys. Rev. B \textbf{87}, 024510 (2013).
\bibitem{R41} Maxime Boissonneault, A. C. Doherty, F. R. Ong, P. Bertet, D. Vion, D. Esteve and A. Blais,
    Back-action of a driven nonlinear resonator on a superconducting qubit, Phys. Rev. A \textbf{85}, 022305 (2012).
\bibitem{R42} Z. H. Wang, C. P. Sun, and Yong Li, Microwave degenerate parametric down-conversion with a single cyclic
three-level system in a circuit-QED setup, Phys. Rev. A \textbf{91}, 043801 (2015).
\bibitem{R43} W. Pfaff, C. J. Axline, L. D. Burkhart, U. Vool, P. Reinhold, L. Frunzio, Liang Jiang, M. H. Devoret, and
R. J. Schoelkopf, Controlled release of multiphoton quantum states from a microwave cavity memory, Nat. Phys. \textbf{13},
882 (2017).
\bibitem{R44} S. Rosenblum, Y. Y. Gao, P. Reinhold, C. Wang, C. Axline, L. Frunzio, S. M. Girvin, L. Jiang, M. Mirrahimi,
M. H. Devoret, and R. J. Schoelkopf, A CNOT gate between multiphoton qubits encoded in two cavities, arXiv:1709.05425 (2017).
\bibitem{R45} G. E. Santoro, R. Marto\v{n}\'{a}k, E. Tosatti, and R. Car, Theory of quantum annealing of an Ising spin
glass, Science \textbf{295}, 2427 (2002).
\bibitem{R46} V. Choi, Minor-embedding in adiabatic quantum computation: I. The parameter setting problem, Quant. Inf.
Proc. \textbf{7}, 193 (2008).
\bibitem{R47} V. Choi, Minor-embedding in adiabatic quantum computation: II. Minor-universal graph design, Quant. Inf.
Proc. \textbf{10}, 343 (2011).
\bibitem{R48} In the LHZ scheme, the logical coupling $\mathcal{J}_{ij}$ is encoded in the local fields applied on the
physical spin, and the logical local fields $h_{i}$ can be treated as a logical coupling $\mathcal{J}_{i0}=h_{i}$, which
descibes the coupling strength between the $i$th logical spin and an ancilla fixed spin (i.e., $|\uparrow\rangle$) labeled by $0$. In this work, for
clarity, we use the notation adopted in Ref.$\,$[36], where $\sigma_{i}^{Z}$ represents Pauli operator for the $i$ physical
spin in the left side diagonal of the triangular lattice, and also the Pauli operator for the $i$ logical spin, as shown
in Fig.$\,$4.
\bibitem{R49} For terms in the base of the triangular lattice, i.e., three-body terms, as shown in Fig.$\,$4(b), by
introducing a fixed physical spin $|\uparrow\rangle$, it can also be realized with four-body terms.
\bibitem{R50} N. Chancellor, S. Zohren, and P. A. Warburton, Circuit design for multi-body interactions in superconducting
quantum annealing system with applications to a scalable architecture, npj Quant. Informa. \textbf{3}, 21 (2017).
\bibitem{R51} A. W. Glaetzle, R. M. W. van Bijnen, P. Zoller, W. Lechner, A coherent quantum annealer with Rydberg atoms.
Nat. Commun. \textbf{8}, 15813 (2017).
\bibitem{R52} H. Goto, Universal quantum computation with a nonlinear oscillator network, Phys. Rev. A \textbf{93}, 050301 (2016).
\bibitem{R53} Z. Xiang, S. Ashhab, J. Q. You, and F. Nori, Hybrid quantum circuits: Superconducting circuits interacting
with other quantum systems, Rev. Mod. Phys. \textbf{85}, 623 (2013).
\bibitem{R54} Z. Kurucz and K. M{\o}lmer, Multilevel Holstein-Primakoff approximation and its application to atomic spin squeezing and ensemble quantum memories, Phys. Rev. A \textbf{81}, 032314 (2010).
\bibitem{R55} T. Opatrn\'{y} and K. M{\o}lmer, Spin squeezing and Schr\"{o}dinger-cat-state generation in atomic samples with Rydberg blockade, Phys. Rev. A \textbf{86}, 023845 (2012).

\bibitem{R56} T. Brecht, W. Pfaff, C. Wang, Y. Chu, L. Frunzio, M. H. Devoret, and R. J. Schoelkopf, Multilayer microwave
integrated quantum circuits for scalable quantum computing, Npj Quant. Inf. \textbf{2}, 16002 (2016).
\bibitem{R57} C. Axline, M. Reagor, R. Heeres, P. Reinhold, C. Wang, K. Shain, W. Pfaff, Y. Chu, L. Frunzio, and R. J.
Schoelkopf, An architecture for integrating planar and 3D cQED devices, Appl. Phys. Lett. \textbf{109}, 042601 (2016).
\bibitem{R58} J. R. Johansson, P. D. Nation, and F. Nori, QuTiP: An open-source Python framework for the dynamics of
open quantum systems, Compu. Phys. Commun. \textbf{183}, 1760 (2012).
\bibitem{R59} J. R. Johansson, P. D. Nation, and F. Nori, QuTiP 2: A Python framework for the dynamics of open quantum
systems, Compu. Phys. Comm. \textbf{184}, 1234 (2013).
\bibitem{R60} J. Q. You, X. Hu, S. Ashhab, and F. Nori, Low-decoherence flux qubit, Phys. Rev. B \textbf{75},
140515(R) (2007).
\bibitem{R61} F. Yan, S. Gustavsson, A. Kamal, J. Birenbaum, A. P. Sears, D. Hover, T. J. Gudmundsen,
              D. Rosenberg, G. Samach, S. Weber, J. L. Yoder, T. P. Orlando, J. Clarke, A. J. Kerman,
              and W. D. Oliver, The flux qubit revisited to enhance coherence and reproducibility, Nat.Commun.
              \textbf{7}, 12964 (2016).

\bibitem{R62} T. Yamamoto, K. Inomata, K. Koshino, P.-M. Billangeon, Y. Nakamura, and J. S. Tsai, Superconducting
flux qubit capacitively coupled to an LC resonator, New J. Phys. \textbf{16}, 015017 (2014).
\bibitem{R63} K. Inomata, T. Yamamoto, P.-M. Billangeon, Y. Nakamura, and J. S. Tsai, Large dispersive shift of cavity
resonance induced by a superconducting flux qubit, Phys. Rev. B \textbf{86}, 140508(R) (2012).
\bibitem{R64} R. Krishnan and J. A. Pople, Approximate fourth-order perturbation theory of the electron correlation
energy, Int. J. Quantum Chem. \textbf{14}, 91 (1978).
\bibitem{R65} H. J. Carmichael, Statistical Methods in Quantum Optics $1$: Master Equations and Fokker-Planck Equations,
  $2$nd ed. (Springer, New York, 2002).



\end{thebibliography}
\end{document}